# New Electron-Doped Superconducting Cuprate $Li_xSr_2CuO_2Br_2$

Tetsuya Kajita[1], Masatsune Kato[1*], Takashi Suzuki[1], Takashi Itoh[2], Takashi Noji[1] and Yoji Koike[1]

*E-mail address: kato@teion.apph.tohoku.ac.jp

[1]*Department of Applied Physics, Graduate School of Engineering, Tohoku University, 6-6-05 Aoba, Aramaki, Aoba-ku, Sendai 980-8579, Japan*

[2]*Center for Interdisciplinary Research, Tohoku University, 6-3 Aoba, Aramaki, Aoba-ku, Sendai 980-8578, Japan*

A new electron-doped superconductor $Li_xSr_2CuO_2Br_2$ with $x$ = 0.15 has successfully been synthesized by an electrochemical Li-intercalation technique. The magnetic susceptibility shows superconductivity of bulk with the superconducting transition temperature $T_c$ = 8 K. This compound is the first electron-doped superconducting cuprate with the $K_2NiF_4$ structure.

Since the discovery of high-$T_c$ superconductivity[1], a large number of superconducting cuprates have been synthesized. Most of them are hole-doped superconductors, including (La, Sr)$_2$CuO$_4$[2] and YBa$_2$Cu$_3$O$_7$.[3]
On the other hand, only two families of electron-doped superconducting cuprates are known: one is T'-(Ln, Ce)$_2$CuO$_4$ (Ln : lanthanide)[4] and the other is the so-called infinite-layer compounds (Sr, Ln)CuO$_2$.[5]

As shown in Fig. 1, Sr$_2$CuO$_2$Br$_2$ is a layered perovskite with the K$_2$NiF$_4$ structure and essentially isostructural to the well-known hole-doped high-$T_c$ superconductor (La, Sr)$_2$CuO$_4$. The Sr$_2$CuO$_2$Br$_2$ contains CuO$_2$ planes as in the case of (La, Sr)$_2$CuO$_4$, but the out-of-plane oxygen ions at the apices of the CuO$_6$ octahedron are replaced by $Br^-$ ions, and (La, Sr) by Sr. In the rock-salt layer of Sr and Br, $Sr^{2+}$ ions shift a little towards the nearest CuO$_2$ plane and $Br^-$ ions away from the CuO$_2$ plane, which is due to the larger radius of $Br^-$ than of $O^{2-}$ and the smaller Coulomb attraction between $Cu^{2+}$ and $Br^-$ than between $Cu^{2+}$ and $O^{2-}$. This leads to the formation of the $Br^-$- $Br^-$ double layers. Accordingly, $Li^+$ ions are expected to be readily intercalated between the electronegative $Br^-$ layers, which are weakly bound through the van der Waals force, as in the case of Li$_x$FeOCl[6] and the superconducting Li$_x$HfNCl (the superconducting transition temperature $T_c$ = 25.5 K).[7]

In this paper, we report the synthesis and superconductivity of a new electron-doped superconducting cuprate, namely, the Li-intercalated layered perovskite Li$_x$Sr$_2$CuO$_2$Br$_2$.

Polycrystalline host samples of Sr$_2$CuO$_2$Br$_2$ were synthesized as follows. First, polycrystals of SrCuO$_2$ were prepared from stoichiometric amounts of SrCO$_3$ and CuO powders. The powders were mixed, ground and heated in air at 925 °C for 10 h. The products were then pulverized, pressed into pellets and sintered for 20 h at 950 °C. Next, the obtained single-phase samples of SrCuO$_2$ were mixed with a stoichiometric amount of SrBr$_2$, pressed into pellets and then sintered for 24 h at 825 °C. Finally, the obtained samples of Sr$_2$CuO$_2$Br$_2$ were mixed with naphthalene of 30 weight %, pelletized with the dimensions of 7 mm in diameter and 1.5 mm in thickness and then sintered again for 6 h at 600 °C to obtain porous samples of Sr$_2$CuO$_2$Br$_2$ which were suitable for the homogeneous intercalation of Li. The electrochemical Li-intercalation was carried out at room temperature in an argon-filled glove box. A three-electrode cell was set up as Sr$_2$CuO$_2$Br$_2$|1.0 M LiClO$_4$/PC|Li. The working electrode was a pellet of Sr$_2$CuO$_2$Br$_2$ which was put between Ni meshes. The counter electrode was a sheet of Li. As an electrolyte, 1.0 M LiClO$_4$ dissolved in propylene carbonate (PC) was used. A sheet of Li was used also as a reference electrode. The Li-intercalation was performed under a constant potential of 0.5 V (vs Li/$Li^+$) using a potentiostat. The total amount of Li intercalated into $Sr_2CuO_2Br_2$ was estimated to be 0.15 according to the simple Faraday law and also the ICP analysis. All products were characterized by powder x-ray diffraction using Cu Kα radiation to be of the single phase. Since the products were unstable in air, they were mixed with grease in order to avoid the exposure to the moisture in the atmosphere during the measurements. The magnetic susceptibility was measured using a SQUID magnetometer in a magnetic field of 3 Oe.

Figure 2 shows the powder x-ray diffraction pattern of $Li_{0.15}Sr_2CuO_2Br_2$. The pattern is the same as that of the host sample. All the peaks can be indexed on the basis of the tetragonal symmetry, indicating no formation of byproducts. Moreover, no change of the lattice parameters through the Li-intercalation is observed within our experimental accuracy.

Figure 3 displays the temperature dependence of the magnetic susceptibility of $Li_{0.15}Sr_2CuO_2Br_2$. A single-step diamagnetic response due to the Meissner effect is observed below 8 K. The superconducting volume fraction estimated at 2 K on field cooling is about 6 %, indicating that the superconductivity is of bulk. The small value may suggest the presence of a pristine region of $Sr_2CuO_2Br_2$ in the sample. At present, we have not yet succeeded in the electrical resistivity measurement because the porous Li-intercalated samples are too brittle for us to make electrically good contact.

The present compound $Li_xSr_2CuO_2Br_2$ is the first electron-doped superconducting cuprate with apical anions. So far, it has been believed that hole carriers cannot be introduced into the $CuO_2$ plane with no apical anion and that electron carriers neither into the $CuO_2$ plane with apical anions. In the $A_2CuO_2X_2$ system (A = alkaline earth metal; X = halogen) with apical halogen ions of X, actually, both $Sr_2CuO_2F_{2+x}$ ($T_c$ = 46 K)[8] and (Ca, Na)$_2CuO_2Cl_2$ ($T_c$ = 26K)[9] are hole-doped superconductors. The noticeable difference in the carrier type is considered to come from the difference in the length of the $a$-axis. The $a$-axis length $a$ = 3.99 Å in the present compound is much larger than those of $Sr_2CuO_2F_2$ ($a$ = 3.81 Å) and $Ca_2CuO_2Cl_2$ ($a$ = 3.84 Å). The increase in the $a$-axis length, namely, the increase in the Cu-O bond length in the $CuO_2$ plane decreases the Madelung potential at the Cu site, so that electronegative electron carriers tend to be readily introduced into the $CuO_2$ plane. According to the empirical rule of the carrier-doping into the $CuO_2$ plane, the critical length of the Cu-O bond in the $CuO_2$ plane at which the carrier type changes from hole-like to electron-like is 1.94 Å, namely, $a$ = 3.88 Å. [10]
The present result is consistent with this rule.

In conclusion, we have successfully synthesized a new superconductor $Li_{0.15}Sr_2CuO_2Br_2$ with $T_c$ = 8 K, which is the third family of electron-doped superconducting cuprates, using the electrochemical Li-intercalation technique. This compound is the first electron-doped superconducting cuprate with apical anions. It has been found that the value of the *a*-axis length is crucial to the carrier type in the superconducting cuprates with the $CuO_2$ plane.

This work was supported by a Grant-in-Aid for Scientific Research from the Ministry of Education, Science, Sports and Culture, Japan.

Figure Captions

Fig. 1 Crystal structure of $Sr_2CuO_2Br_2$.

Fig. 2 Powder x-ray diffraction pattern of $Li_{0.15}Sr_2CuO_2Br_2$ at room temperature. The hump around 20° is due to grease mixed with the powdered sample in order to avoid the exposure to the moisture in the atmosphere.

Fig. 3 Temperature dependence of the magnetic susceptibility of $Li_{0.15}Sr_2CuO_2Br_2$ measured in a magnetic field of 3 Oe on warming after zero-field cooling and then on cooling in a field. The inset shows the onset of the superconducting transition on an expanded scale.

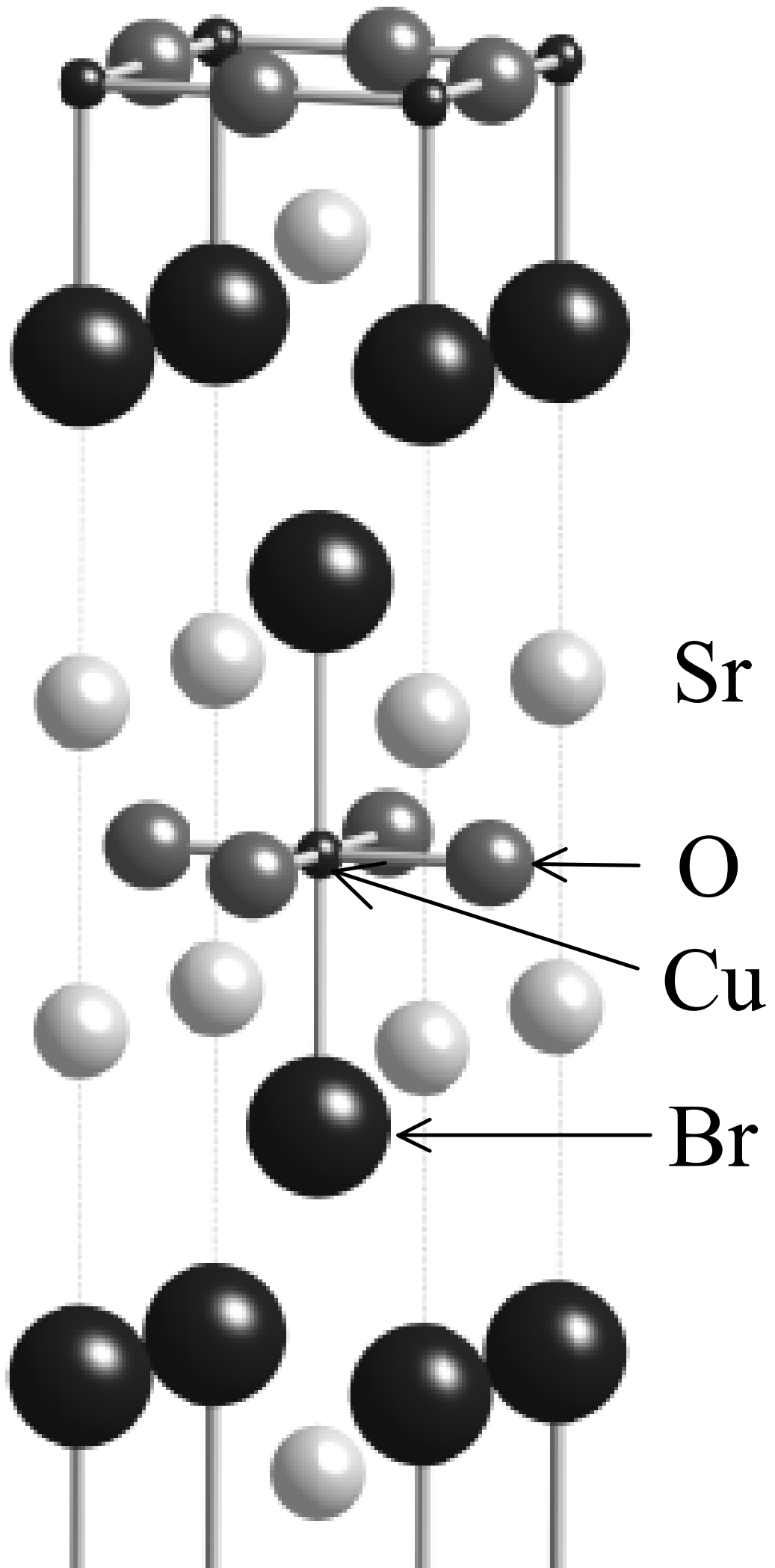

Fig. 1

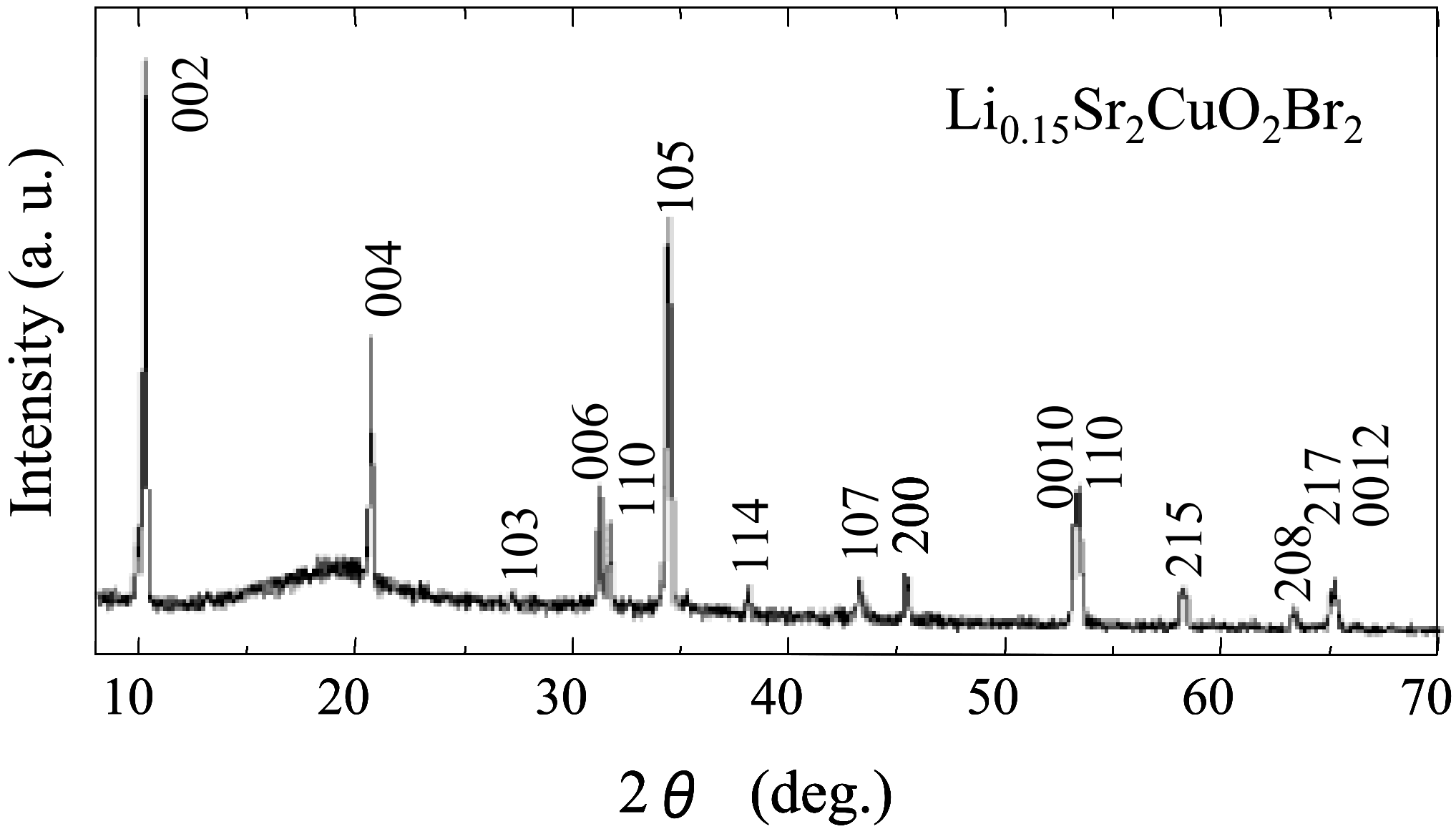

Fig. 2

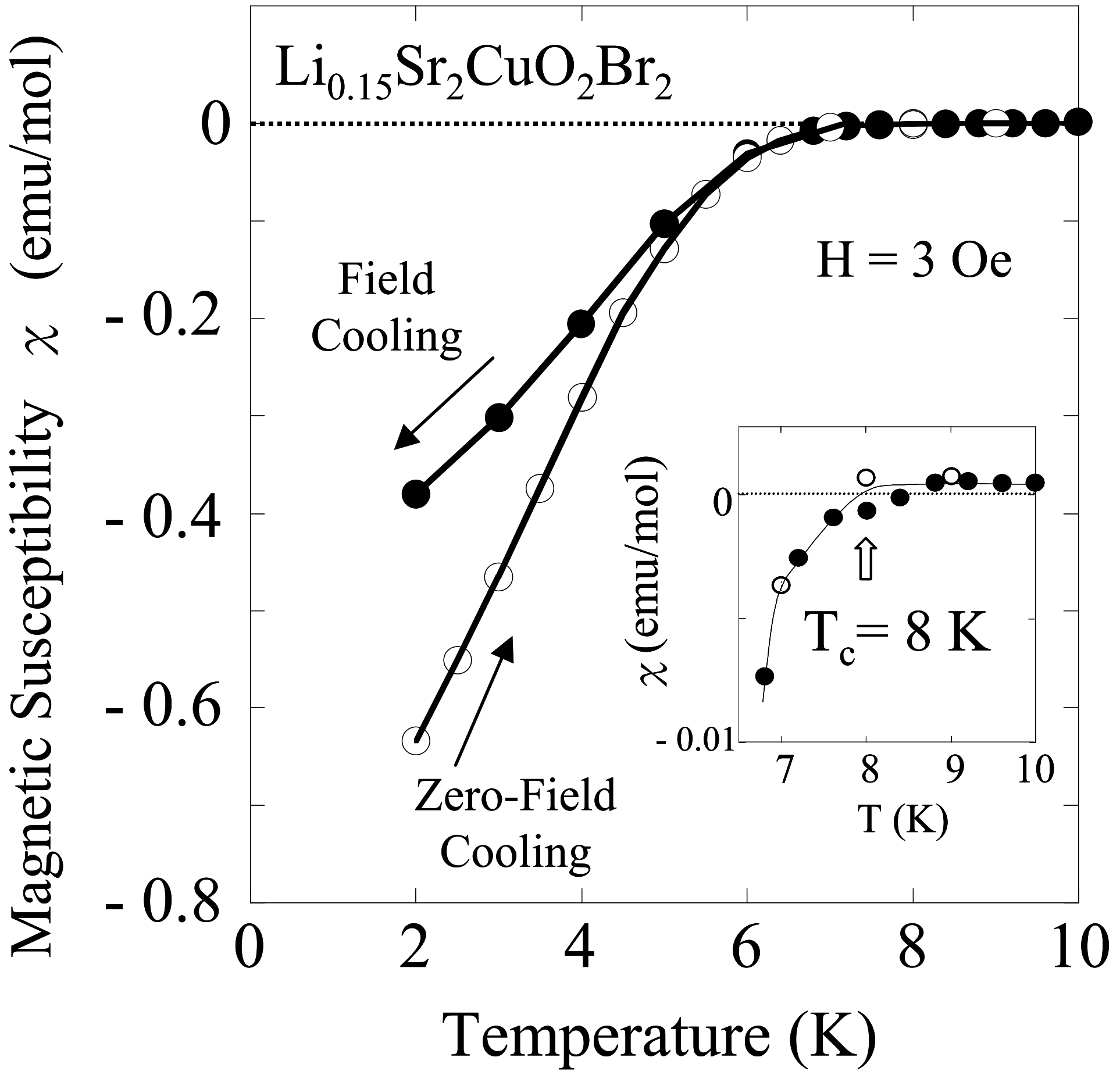

Fig. 3